\documentclass[twocolumn,showpacs,preprintnumbers,amsmath,amssymb]{revtex4}
\usepackage{graphicx}
\usepackage{dcolumn}
\usepackage{bm}
\usepackage{epsfig}

\begin{document}

\title{Measurement of the Nucleon $F^n_2/F^p_2$ Structure Function Ratio 
 by the Jefferson Lab MARATHON Tritium/Helium-3 Deep Inelastic Scattering Experiment}

\author
{ 
 {D.~Abrams},$^{1}$ 
 {H.~Albataineh},$^{2}$
 {B.~S.~Aljawrneh},$^{3}$
 {S.~Alsalmi},$^{4,5}$
 {K.~Aniol},$^{7}$
 {W.~Armstrong},$^{8}$
 {J.~Arrington},$^{8,9}$
 {H.~Atac},$^{10}$
 {T.~Averett},$^{11}$
 {C.~Ayerbe Gayoso
% ...
} $^{26}$
}
\vspace*{0.3cm}
\affiliation{$^{1}$University of Virginia, Charlottesville, VA 22904, USA}
\affiliation{$^{2}$Texas A \& M University, Kingsville, TX 78363, USA}
\affiliation{$^{3}$North Carolina A \& T State University, Greensboro, NC 27411, USA}
\affiliation{$^{4}$Kent State University, Kent, OH 44240, USA}
\affiliation{$^{5}$King Saud University, Riyadh 11451, Kingdom of Saudi Arabia}
% ...
\affiliation{$^{7}$California State University, Los Angeles, CA 90032, USA}
\affiliation{$^{8}$Argonne National Laboratory, Lemont, IL 60439, USA}
\affiliation{$^{9}$Lawrence Berkeley National Laboratory, Berkeley, CA 94720, USA}
\affiliation{$^{10}$Temple University, Philadelphia, PA 19122, USA}
\affiliation{$^{11}$William \& Mary, Williamsburg, VA 23187, USA}
% ...
\affiliation{$^{26}$Stony Brook, State University of New York, NY 11794, USA}
% ...


\date{\today}

\begin{abstract}

\hspace*{1.17in} The Jefferson Lab Hall A Tritium Collaboration \\


The ratio of the nucleon $F_2$ structure functions, $F_2^n/F_2^p$, is determined by the MARATHON
experiment from measurements of deep inelastic scattering of electrons from $^3$H and $^3$He nuclei.
The experiment was performed in the Hall A Facility of Jefferson Lab and used two high resolution
spectrometers for electron detection, and a cryogenic target system which included a low-activity
tritium cell.  The data analysis used a novel technique exploiting the mirror symmetry of the two
nuclei, which essentially eliminates many theoretical uncertainties in the extraction of the ratio.
The results, which cover the Bjorken scaling variable range $0.19 < x < 0.83$, represent a significant
improvement compared to previous SLAC and Jefferson Lab measurements for the ratio.  They are
compared to recent theoretical calculations and empirical determinations of the $F_2^n/F_2^p$ ratio.

\end{abstract}


\maketitle   

The nucleon structure functions, found from deep inelastic scattering (DIS) of electrons by protons
and deuterons, have been of fundamental importance in establishing the internal quark structure of the
nucleon~\cite{fr91}. First measurements occurred in a series of DIS experiments at the Stanford Linear
Accelerator Center (SLAC) circa 1970~\cite{fr72}, which showed the existence of pointlike entities
within the nucleons.  Further studies of muon-nucleon and neutrino-nucleon DIS experiments at
CERN~\cite{au87,be90,ar97,be91} and Fermilab~\cite{ad95,ol92} established the quark-parton model~(QPM)
for the nucleon~\cite{cl79}, and provided supporting evidence for the emerging theory of quantum
chromodynamics~(QCD).

The cross section for deep inelastic electron-nucleon scattering, where the nucleon breaks up, is given,
in the one-photon-exchange approximation, in terms of the structure functions $F_1(\nu,Q^2)$ and
$F_2(\nu,Q^2)$ of the nucleon.  In the lab frame and in natural units it reads~\cite{cl79}:
\begin{equation}
\label{eqsig}
\vspace* {-.12in}
{ 
\vspace* {-.05in}
{ {d^2\sigma} \over {d\Omega dE'} } = \sigma_{M}
\left[ { F_2(\nu,Q^2) \over \nu } + 
 { {2 F_1(\nu,Q^2)} \over {M} } \tan^2 \left( {\theta \over 2} \right) \right]
},
\end{equation}
where $\sigma_{M}={ {4 \alpha^2 (E')^2} \over {Q^4} }  \cos^2 \left( {\theta \over 2} \right)$ is the
Mott cross section, $\alpha$ is the fine-structure constant, $E$ is the incident electron energy,
$E'$ and $\theta$ are the scattered electron energy and angle, $\nu = E-E'$ is the energy transfer,
$Q^2=4 E E' \sin^2(\theta/2)$ is the negative of the four-momentum transfer squared, and $M$ is the
nucleon mass.  The invariant mass of the final hadronic state is $W=(M^2+2M\nu-Q^2)^{1/2}$.

The scattering process is mediated through the exchange of virtual photons. The cross section
can also be written in terms of those for the absorption by the nucleon of longitudinally,
$\sigma_L$, or transversely, $\sigma_T$, polarized photons.  The functions $F_1$ and $F_2$
are related to the ratio $R=\sigma_L/\sigma_T$ as $F_1=MF_2(\nu^2+Q^2)/[Q^2 \nu (1+R)]$~\cite{fr72}.
All of the above formalism can also be applied to the case of DIS by a nucleus, with $F_1$ and
$F_2$ becoming the structure functions of the nucleus in question.  It should be noted that the
ratio of DIS cross sections of different nuclear targets is equivalent to the ratio of their $F_2$
structure functions if $R$ is the same for all nuclei.  The latter has been confirmed experimentally
within inherent experimental uncertainties~\cite{ta96}.  

The basic idea of the QPM~\cite{bj69,fe69} is to represent DIS as quasi-free scattering of electrons
from the nucleon's partons/quarks, in a frame where the nucleon possesses infinite momentum. The
fractional momentum of the nucleon carried by the struck quark is then given by the Bjorken ``scaling"
variable, $x=Q^2/(2M\nu)$. In the limit where $\nu \rightarrow \infty$, $Q^2 \rightarrow \infty$ with
$x$ finite between 0 and 1, the nucleon structure functions become: 
$F_1 = {1 \over 2} \sum_i e_i^2 f_i(x)$ and $F_2 = x \sum_i e_i^2f_i(x)$,
where $e_i$ is the fractional charge of quark type $i$, $f_i(x)dx$ is the probability that a quark of
type $i$ carries momentum in the range between $x$ and $x+dx$, and the sum runs over all quark types.
For the Gell-Mann/Zweig quarks, the $F_2(x)$ structure function for the proton ($p$) becomes                 
$F^p_2(x) = x [(4/9)U+(1/9)D+(1/9)S]$, and due to isospin symmetry, that of the neutron ($n$)
$F^n_2(x) = x \left[(1/9) U + (4/9) D + (1/9) S \right]$, where $U=u+\bar{u}$, $D=d+\bar{d}$, and
$S=s+\bar{s}$, with bars denoting antiquarks~\cite{cl79}.

The positivity of the structure functions dictates that the ratio of the neutron to proton $F_2$ 
functions is bounded for all values of $x$: $(1/4) \leq {F^n_2/F^p_2 } \leq 4$, a relationship known
as the Nachtmann inequality~\cite{na72}.  This relationship was verified in the pioneering SLAC
experiments E49a and E49b circa 1970~\cite{bo73}, which found that the ratio approaches unity at $x=0$
and approximately 1/4 at $x=1$.  The SLAC findings showed that at low $x$ the three quark-antiquark
distributions are equal, dominated by sea quarks, and that at large $x$ the $u$~($d$) quark distribution
dominates in the proton~(neutron).  These findings were surprising as the expectation, at the time,
from SU(6) symmetry was that $F_2^n/F_2^p$ should be equal to 2/3 for all $x$.  The behavior of the
ratio at $x=1$ was justified by the diquark model of Close~\cite{cl73}, and Regge phenomenology, 
initiated by Feynman~\cite{fe75}.  In Close's diquark model, the diquark configuration with
spin 1 is suppressed relative to that with spin 0.  The phenomenological suppression of the $d$
quark distribution, which results from the $F_2^n/F_2^p$ value of 1/4 at $x=1$, can be understood
in the quark model of  Isgur~\cite{is99} in terms of the color-magnetic hyperfine interaction
between quarks, which is also responsible for the $N$-$\Delta$ mass splitting.  It should be noted
that perturbative QCD arguments~\cite{fa75} and a treatment based on quark-counting rules~\cite{br95}
suggest that the nucleon $F_2$ ratio should have the larger value of 3/7 at $x=1$.

The original considerations of the magnitude of the nucleon $F_2$ ratio were called into question in the
1990s when a re-examination of the subject by Whitlow {\it et al.}~\cite{wh92}, who, using the original
SLAC data~\cite{bo73} and a plausible model of the EMC effect in which the deuteron, medium and heavy
nuclei scale with nuclear density~\cite{fr88}, found a strong sensitivity in the determination of the ratio 
at large $x$.  The EMC effect, discovered at CERN~\cite{au83} and quantified precisely at SLAC~\cite{go94},
characterizes the modification of the nucleon structure functions in nuclear matter.  The above strong
sensitivity was subsequently confirmed in a relativistic re-analysis of the SLAC data, which assumes the
presence of minimal binding effects in the deuteron~\cite{me96}.  In Ref.~\cite{wh92},
it also became evident that the nucleon $F_2$ ratio was very sensitive to the choice of the nucleon-nucleon
(N-N) potential model governing the structure of deuterium, later confirmed in Refs.~\cite{ac11,ar12}. The
large uncertainty in the extraction of the $F^n_2/F^p_2$ ratio at large $x$ calls into question the
presumption that $F^n_2/F^p_2$ and $D/U$ tend to 1/4 and zero, respectively, as $x$ approaches 1.

These difficulties in the $F^n_2/F^p_2$ determination can be remedied using a method proposed by
Afnan {\it et al.}~\cite{af00,af03}, which determines the $F_2^n/F_2^p$ ratio from DIS measurements on
$^3$H (triton) and $^3$He (helion), exploiting the isospin symmetry and similarities of the two $A=3$
mirror nuclei.  In the absence of Coulomb interactions and for an isospin symmetric world, the properties
of a proton (neutron) bound in the $^3$He nucleus should be identical to that of a neutron (proton) bound
in the $^3$H nucleus.  Defining the EMC-type ratios for the $F_2$ structure functions of helion ($h$) and
triton ($t$) by: $R_h = F_2^h/(2 F_2^p + F_2^n)$ and $R_t = F_2^t/(F_2^p + 2 F_2^n)$, one can write the
ratio of these ratios as ${\cal R}_{ht} = R_h/R_t$, which directly yields the $F_2^n/F_2^p$ ratio as: 
\begin{equation} 
\label{super} 
{ F_2^n \over F_2^p } 
= { 2 {\cal R}_{ht} - F_2^h/F_2^t 
\over 2 F_2^h/F_2^t - {\cal R}_{ht} } . 
\end{equation} 
The $F_2^n/F_2^p$ ratio found from this Equation depends on the ratio of the EMC effects in $^3$He and $^3$H.
Since the neutron and proton distributions in the $A=3$ nuclei are similar, the ratio can be calculated
reliably with the expectation that ${\cal R}_{ht} \simeq 1$~\cite{af03,pa01}, once $F_2^h/F_2^t$ is
measured experimentally.  The seeming dependence of the process on the $F_2^n/F_2^p$ input is actually
artificial.  In practice, one can employ an iterative procedure to eliminate this dependence altogether.
Namely, after extracting $F_2^n/F_2^p$ from the data using some calculated ${\cal R}_{ht}$, the extracted
$F_2^n/F_2^p$ can then be used to compute a new ${\cal R}_{ht}$, which is then used to extract a new and better
value of $F_2^n/F_2^p$. This procedure is iterated until convergence is achieved and a self-consistent solution
for the extracted $F_2^n/F_2^p$ is obtained. The convergence of the procedure was confirmed in
Refs.~\cite{pe10,pa01}.
  
The above technique was used in the JLab E12-10-103 MARATHON experiment~\cite{pe10} (initiated in 1999~\cite{pe99}),
which took data in the winter/spring of 2018 using the Electron Accelerator and Hall A Facilities of the Lab.
Electrons scattered from light nuclei were detected in the Left and Right High Resolution Spectrometers (LHRS
and RHRS) of the Hall~\cite{al04}.  The beam energy was fixed at 10.59 GeV, and its current ranged from 14.6 to
22.5~$\mu$A.  The experiment detected DIS events from the proton, deuteron, helion, and triton particles using
a cryogenic gaseous target system~\cite{ho10}.  The LHRS was operated at a momentum of 3.1 GeV/{\it c} with
angles between $16.81^{\circ}$ and $33.55^{\circ}$.  The RHRS was operated at a single setting of 2.9 GeV/{\it c}
and $36.12^{\circ}$.     

The target system consisted of four high-pressure cells, of length 25.0~cm and diameter 1.27~cm, containing
$^3$He, $^3$H, $^2$H, and $^1$H gases.  The four cells were filled at temperatures of 294.3, 296.3, 296.1,
297.4 K, and pressures of 17.19, 13.82, 35.02, 35.02 atm, resulting in densities (determined from
data-supported virial models~\cite{ga12}) of $2.129\pm0.021$, $3.400\pm0.010$, $5.686\pm0.022$, and
$2.832\pm0.011$ kg/m$^3$, respectively. The target assembly also contained an empty cell
and a ``dummy target" consisting of two Al foils separated by 25.0~cm, which were used to measure the
contribution to the scattered electron yields from the Al end-caps of the cells.  The cells were
cycled many times in the beam for each kinematic setting in order to minimize effects of possible drifts of
the beam diagnostic or other instrumentation ({\it e.g.} the beam current monitors).  

Scattered particles were detected in the HRSs using two planes of scintillators for event triggering,
two drift chambers for particle tracking, and a gas threshold \v{C}erenkov counter and a lead-glass calorimeter
for particle identification.  Electrons were identified as electrons on the basis of i) a minimal pulse height
in the \v{C}erenkov counter, and ii) the energy deposited in the calorimeter, consistent with the momentum as
determined from the drift chamber track using the spectrometers' optical properties.  The detector
efficiencies for both spectrometers were found to be stable and independent of
the gas target used.  A small fraction of events with two or more drift chamber tracks
(1-2\% of the total) were not included in the data analysis, as they were dominated by electrons passing through
the edges of the exit of the Al vacuum pipe of the spectrometers.  Details on the Hall A Facility, beam line,
and detector instrumentation as used in MARATHON, including calibrations, are given in 
Refs.~\cite{ba19,ha20,ku19,li20,ny20,su20}. 

Because of the low density of the gas targets, the electron counting rate was dominated, for all kinematics,
by events originating from the target cell Al end-caps, as the total thickness of the two end-caps of the
$^3$He, $^3$H, $^2$H and $^1$H cells was 0.55, 0.60, 0.51, and 0.64 mm, respectively.  In order to reject
electrons originating from the end-caps, a software, scattering-angle-dependent target
position reconstruction ``cut" was imposed, which resulted in an effective, usable target length of 21 cm, on
the average.

All events properly identified as electrons originating from the gas inside each target cell were binned by
Bjorken $x$, resulting in the formation of an electron yield, $Y(x)$, defined as:
\begin{equation}
Y(x) = { { {N_{{\rm e}'}} \over {N_{\rm e} \rho_t L_t } } C_{\rm cor} },
\end{equation}
where $N_{{\rm e}'}$ is the number of scattered electrons, $N_{\rm e}$ is the number of incident beam electrons,
$\rho_t$ is the density of the gas target, $L_t$ is the selected target length, and 
$C_{\rm cor}=C_{\rm det}C_{\rm cdt}C_{\rm den}C_{\rm tec}C_{\rm psp}C_{\rm rad}C_{\rm cde}C_{\rm bin}C_{\rm dth}$.
Here, $C_{\rm det}$ is the correction for trigger and detector inefficiency, $C_{\rm cdt}$ is the computer dead-time
correction (1.001 to 1.065), $C_{\rm den}$ is a correction to the target density due to beam heating effects
(1.066 to 1.125), $C_{\rm tec}$ is a correction for falsely-reconstructed events originating from the end-caps
(0.973 to 0.998), $C_{\rm psp}$ is a correction for events originating from pair symmetric processes (0.986
to 0.999), $C_{\rm rad}$ is a correction for radiative effects (0.826 to 1.173), $C_{\rm cde}$ is a correction for
Coulomb distortion effects (0.997 to 1.000), $C_{\rm bin}$ is a bin-centering correction (0.995 to 1.001), and 
$C_{\rm dth}$ is a correction for the beta decay of tritons to helions, applicable only to the tritium yield
[0.997~(0.989) at the beginning~(end) of the experiment].  A cross section model by Kulagin and Petti (K-P), based
on the works of Refs.~\cite{al17,ku10,ku06}, was adopted~\cite{kppc} for the bin-centering correction, and the
Coulomb correction (which used the $Q^2$-effective approximation as outlined in Ref.~\cite{ub71}).

\begin{figure} [ht!]
\includegraphics[width=\columnwidth, angle=0]{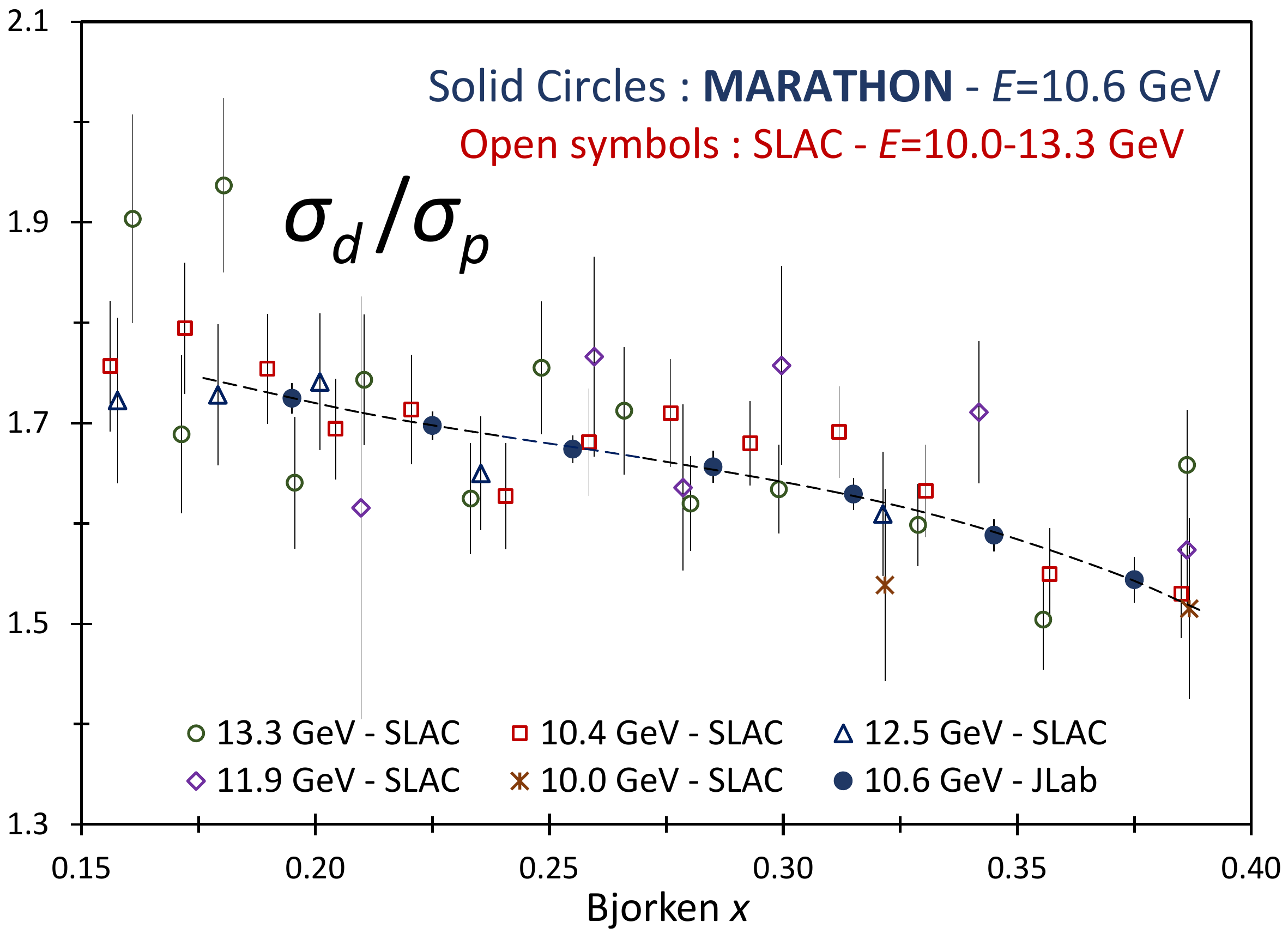}
\caption{\label{figdp} The ratio of the DIS cross sections of deuteron and proton plotted versus the
Bjorken $x$ from the JLab Hall A MARATHON experiment.  Also shown are seminal SLAC data~\cite{bo79} in
the same kinematic region as MARATHON (see text). The dashed curve is a fit to the MARATHON data.  The
MARATHON error bars include all uncertainties added in quadrature.  The SLAC error bars are dominated
by statistical uncertainties, and do not include an overall normalization uncertainty of $\pm 1.3\%$.}
\end{figure}

When forming ratios of electron yields from different targets, which are equivalent to cross section
ratios, the effective gas target length $L_t$ (18.0-22.5 cm) and the correction $C_{\rm det}$ cancel
out. In general, the corrections to the {\it ratios} from each effect become minimal, and in some cases, 
so do the associated systematic uncertainties. For example, the radiative effect correction, ranges
from 0.997 (at the highest $x$) to 1.015 (at the lowest $x$) for the $h/t$ cross section ratio. The
dominant point-to-point systematic uncertainties for the yield ratios are those from the beam-heating
gas target density changes [$\pm(0.1\%$-$0.5\%)$], the radiative correction [$\pm(0.25\%$-$0.45\%)$],
and the choice of spectrometer acceptance limits ($\pm0.2\%$).  The total point-to-point uncertainty
ranged from $\pm0.4\%$ to $\pm1.0\%$ for the $d/p$ cross section ratio, and $\pm0.3\%$ to $\pm0.5\%$
for the $h/t$ ratio.  Details on the determination of the yields, and all associated corrections and
uncertainties, can be found in Refs.~\cite{ha20,ku19,li20,ny20,su20}.

The experiment also collected DIS data for the proton and deuteron~($d$) over the $x$ range from 0.19
to 0.37 for the purpose of finding the $F_2^n/F_2^p$ ratio in the vicinity of $x=0.3$, where it is known
that nuclear corrections are minimal~\cite{al17,ku06}, and comparing it with the $F_2^n/F_2^p$ ratio found
using DIS by the triton and helion.  The measured values of the $\sigma_{d}/\sigma_p$ ratio are given, together
with associated uncertainties, in Table 1 of the Appendix.
The $\sigma_{d}/\sigma_p=F_2^{d}/F_2^p$ values, plotted in Figure \ref{figdp},
are compared to reference measurements from the seminal SLAC E49b and E87 experiments~\cite{bo79}
taken with similar beam energies.  It is evident from Fig.~\ref{figdp} that the JLab and SLAC data are in
excellent agreement, at the $10^{-3}$ level.  Given the ratio ${\cal R}_d=F_2^{d}/(F_2^p+F_2^n)$,
the $F_2^n/F_2^p$ ratio is calculated as $F_2^n/F_2^p=(F_2^d/F_2^p)/{\cal R}_d-1$~\cite{pa01,ku10}.  The
${\cal R}_d$ ratio used in the MARATHON $F_2^{d}/F_2^p$ data analysis is from the model by Kulagin and Petti
based on Refs.~\cite{al17,ku10}.  The results of this model are, in the vicinity of $x=0.3$, in excellent
agreement with determinations using data from the JLab BoNuS~\cite{gr15} and SLAC E139~\cite{go94} experiments,
and two distinct calculations based on studies of data from DIS off nuclei, described in 
Refs.~\cite{ku06} (using nuclei with $A \geq 4$) and~\cite{se20} (using nuclei with $A \geq 3$).

\begin{figure} [t!]
\includegraphics[width=\columnwidth, angle=0]{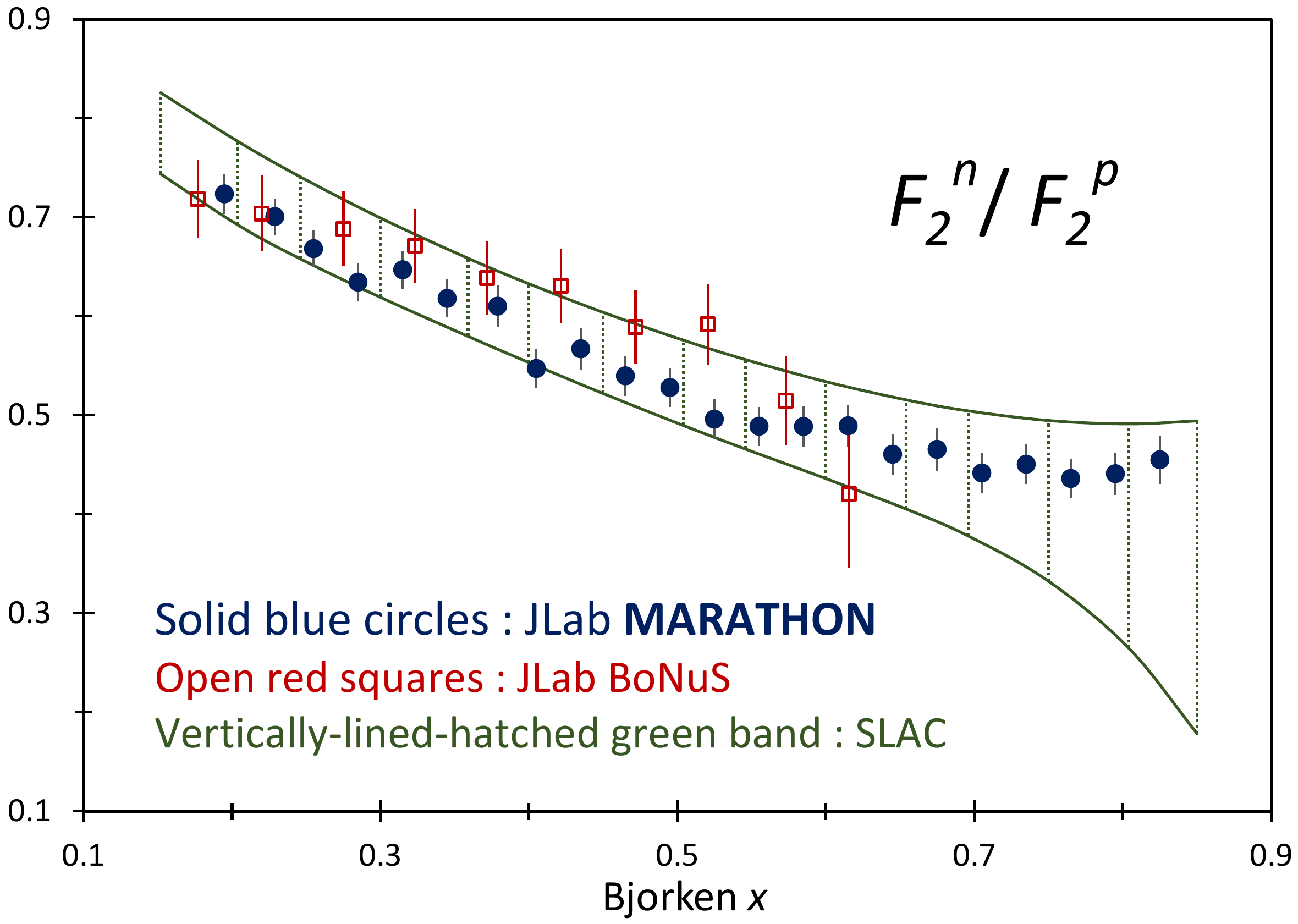}
\caption{\label{fignp} The $F_2^n/F_2^p $ ratio plotted versus the Bjorken $x$ from the JLab MARATHON
experiment.  Also shown are JLab Hall B BoNuS data~\cite{tk14}, and a band based on the fit
of the SLAC data as provided in Ref.~\cite{bo79}, for the MARATHON kinematics [$Q^2=14 \cdot x$
(GeV/{\it c})$^2$] (see text).  All three experimental data sets include statistical, point to point systematic,
and normalization uncertainties.}
\end{figure}

The focus of MARATHON was to study DIS from helion and triton in order to extract the $F_2^n/F_2^p$ ratio in
the range $0.19<x<0.83$ using the measured $\sigma_h/\sigma_t=F_2^h/F_2^t$ ratio and model-calculated values
of the super-ratio ${\cal R}_{ht}$.  The values used for ${\cal R}_{ht}$ come from the theoretical model by
Kulagin and Petti~\cite{ku10,ku06}, which provides a global description of the EMC effect for all known targets
(for a review see Ref.~\cite{ku16}).  The K-P model includes a number of nuclear effects out of which the major
correction for the relevant kinematics comes from the smearing effect with the nuclear energy-momentum distribution,
described in terms of the nuclear spectral function, together with an off-shell correction to the bound nucleon
$F_2$~\cite{ku06}.  The underlying nucleon structure functions come from the global QCD analysis of
Ref.~\cite{al07}, which was performed up to NNLO approximation in the strong coupling constant
including target mass corrections~\cite{ge76} as well as those due to higher-twist effects.
For the spectral functions of the ${^3}$H and ${^3}$He nuclei, the results of Ref.~\cite{pa01} have been used.
In order to evaluate theoretical uncertainties, the ${^3}$He spectral function of Ref.~\cite{sc92} was used.
Reasonable variations of the high-momentum part of the nucleon momentum distribution in ${^3}$H and ${^3}$He were
considered, and uncertainties in the off-shell correction of Ref.~\cite{ku06}, as well as in the nucleon structure
functions, were accounted for.  The maximum resulting uncertainty in ${\cal R}_{ht}$ is estimated to be up to
$\pm 0.4\%$ (at $x=0.8$), contributing minimally to the total uncertainty in the final $F_2^n/F_2^p$ values.
The K-P calculations were performed prior to the analysis of the MARATHON data.

The comparison of $F_2^n/F_2^p$ as extracted from $\sigma_h/\sigma_t$ and $\sigma_d/\sigma_p$ was done at
$x=0.31$, where nuclear corrections contribute negligibly to EMC-type ratios like ${\cal R}_d$ and
${\cal R}_{ht}$, as $\sigma_A/A=\sigma_d/2$~\cite{we11} (determined by the $A \geq 3$ data of
Refs. \cite{go94,ai03,se09} and taking into account the quoted normalization uncertainties therein).  The K-P
models used, predicted a value of 1.000 at $x=0.31$ for both ${\cal R}_{ht}$ and ${\cal R}_d$ with uncertainties
of $\pm0.1\%$ and $\pm0.2\%$, respectively.  The recent work of Ref.~\cite{se20}, based on a global analysis of
nuclear DIS data where the EMC effect is accounted for through nucleon short-range correlations, found 
${\cal R}_{ht}(x=0.31)=1.001$, with a similar uncertainty.  The values of $\sigma_{d}/\sigma_p$ and 
$\sigma_h/\sigma_t$ at $x=0.31$ were determined by
weighted fits to the MARATHON data, which included statistical and point-to-point uncertainties added in
quadrature.  In order to match the $F_2^n/F_2^p$ values found using the two different sets of nuclei, the 
$\sigma_h/\sigma_t$ ratio at $x=0.31$ had to be normalized by a multiplicative factor of 1.025$\pm$0.007.
Consequently, all values of $\sigma_h/\sigma_t$ reported in this work have been normalized upwards by $2.5\%$.

The normalized $\sigma_h/\sigma_t$ values are given in Table 2 of the Appendix,
together with associated uncertainties.  The $F_2^n/F_2^p$ values are given in Table 3 of the Appendix,
together with associated uncertainties.  Shown also in Table 3 are the ${\cal R}_{ht}$ super-ratio values
used to find $F_2^n/F_2^p$.  The ${\cal R}_{ht}$ uncertainty was incorporated in quadrature with the
point-to-point $F_2^n/F_2^p$ uncertainty.  Figure \ref{fignp} shows the MARATHON results for the $F_2^n/F_2^p$
ratio, along with data from the JLab Hall B BoNuS experiment~\cite{tk14} for $W \geq 1.84~{\rm GeV}/{\it c}^2$,
evolved to the $Q^2$ of MARATHON~\cite{wh92}, and results from early SLAC measurements with
$W \geq 1.84~{\rm GeV}/{\it c}^2$~\cite{bo73,bo79}. The SLAC results are presented
as a band, the width of which at high $x$ is dominated primarily by uncertainties due the choice of the
N-N potential used for the evaluation of the deuteron wave function~\cite{wh92,ac11,ar12}.
The MARATHON data are in good agreement with the BoNuS data, and fall well within the SLAC results band. 
The highest-$x$ points are consistent with the $F_2^n/F_2^p$ ratio tending to a value between 0.4 and
0.5 at $x=1$.  This is consistent with the predictions of perturbative QCD and quark counting rules (for
which this ratio is 3/7 at $x=1$), and with a recent prediction~\cite{ro13} that treats strong interactions
using the Dyson-Schwinger equations, where diquark correlations in the nucleons are consequences of
dynamical chiral symmetry breaking (for which the nucleon $F_2$ ratio lies, at $x=1$, between 0.4 and 0.5). 
It is also consistent with a covariant quark-diquark model which also predicts that this ratio should 
be 3/7 at $x=1$~\cite{cl05}.  

\begin{figure} [t!]
\includegraphics[width=\columnwidth, angle=0]{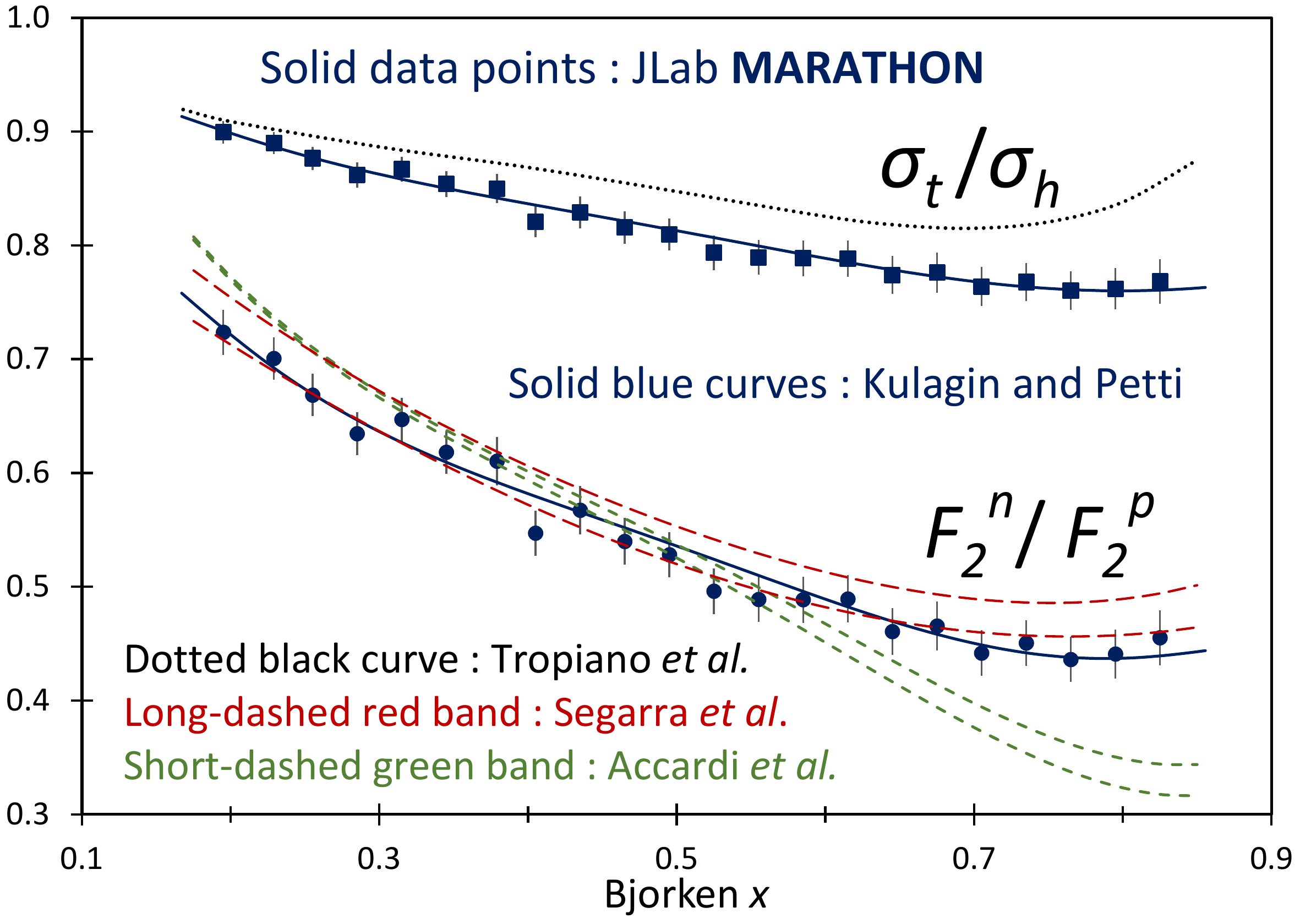}
\caption{\label{figkp} The DIS $\sigma_t/\sigma_h$ and the $F_2^n/F_2^p$ ratios from the MARATHON experiment,
plotted versus the Bjorken $x$, compared to the theoretical predictions of Kulagin and Petti, and of 
Refs.~\cite{se20,ac16,tr19} (see text).  The error bars include overall systematic uncertainties.  All curves
correspond to the MARATHON kinematics, except for the dotted curve, calculated at a fixed $Q^2$ of
4~(GeV/{\it c})$^2$.}
\end{figure}

The MARATHON $F_2^n/F_2^p$ ratio values are in excellent agreement, as quantified by a $\chi^2$ per degree
of freedom (df) of 0.8, with those predicted by Kulagin and Petti, which were used in the determination of
${\cal R}_{ht}$.  For this reason, an iterative procedure, as described earlier, was not necessary.  A
comparison between the MARATHON $F_2^n/F_2^p$ results and the K-P prediction is shown in Figure \ref{figkp}.
Shown also in the Figure are the $\sigma_t/\sigma_h$ MARATHON values compared with the K-P prediction.
The predicted $\sigma_t/\sigma_h$ values by K-P, which were also used in the determination of ${\cal R}_{ht}$,
are in excellent agreement with the MARATHON data, as quantified by a $\chi^2$/df of 0.8.  Also shown in
Figure \ref{figkp} is the nuclear DIS determination of $F_2^n/F_2^p$ by Segarra {\it et al.}~\cite{se20}, the
latest calculation for $F_2^n/F_2^p$ by the CTEQ-JLab~(CJ) Collaboration~\cite{ac16}, and a recent prediction
for $\sigma_t/\sigma_h$ by Tropiano {\it et al.}~\cite{tr19}, which includes isovector components in the
off-shell effects for the bound nucleons in the two $A=3$ nuclei, resulting in different corrections for the
proton and neutron.

In summary, the JLab MARATHON experiment has provided a precise determination of the nucleon $F_2^n/F_2^p$
ratio, which is expected to constrain theoretical models of the few body nuclear structure functions, and
to be used in algorithms which fit~\cite{ac16,sa16,al17} hadronic data to determine the essentially unknown
$(u+\bar{u})/(d+\bar{d})$ ratio at large Bjorken $x$.  These new data will also improve our knowledge of the
nucleon parton distributions, which is needed for the interpretation of high-energy collider data.        

We acknowledge the outstanding support of the staff of the Accelerator Division and Hall A Facility of
JLab, and work of the staff of the Savannah River Tritium Enterprises and the JLab Target Group.
We thank Dr. M.~E.~Christy for useful discussions on the optical properties of the HRS systems.
We are grateful to Dr. W.~Melnitchouk for his contributions to the development of the MARATHON proposal, 
and to Dr. A.~W.~Thomas for many valuable discussions on and support of the MARATHON project 
since its inception.
This material is based upon work supported by the U.S. Department of Energy (DOE), Office of Science,
Office of Nuclear Physics under contracts DE-AC05-06OR23177 and DE-AC02-06CH11357.
This work was also supported by DOE contract DE-AC02-05CH11231, DOE award DE-SC0016577,
National Science Foundation awards NSF-PHY-1405814 and NSF-PHY-1714809, the 
Kent State University Research Council, the Pazy Foundation and the Israeli Science Foundation under grants
136/12 and 1334/16, and the Italian Institute of Nuclear Physics.

\newpage
\onecolumngrid

\centerline{\bf Appendix - Tables of Measurements}
\centerline{D. Abrams {\it et al.} (The Jefferson Lab Hall A Tritium Collaboration)} 
\centerline{\it Measurement of the Nucleon F2n/F2p Structure Function Ratio by the Jefferson}
\centerline{\it Lab MARATHON Tritium/Helium-3 Deep Inelastic Scattering Experiment}
   
\begin{table} [ht!]
\begin{center}
\begin{tabular}{cccccccc}
\hline\hline
      &   &  &                     &                 &       &         &                \\
$x$   & $Q^2$	 & $W$ & $\sigma_{d}/\sigma_p$ & $\Delta_{\rm stat}$ & $\Delta_{\rm ptp}$ & $\Delta_{\rm syst}$ & $\Delta_{\rm tot}$ \\
      & $({\rm GeV}/c)^2$ & ${\rm GeV}/c^2$ &                &             &       &         &           \\
      &   &  &                     &                 &       &         &                \\
\hline\hline
      &   &  &                     &                 &       &         &                \\
0.195 &	2.73     &	3.49	         &      1.725          & 0.005	         &	0.010 	  & 0.009 &	0.015 	\\
0.225 &	3.15	 &	3.42	         &	1.697          & 0.005	         &	0.008 	  & 0.009 &	0.014 	\\
0.255 &	3.57	 &	3.36         	 &	1.674	       & 0.007	         &	0.008 	  & 0.009 &	0.014 	\\
0.285 &	3.99	 &	3.30	         &	1.656	       & 0.008	         &	0.010 	  & 0.009 &	0.016 	\\
0.315 &	4.41	 &	3.24	         &	1.629	       & 0.008	         &	0.010 	  & 0.009 &	0.016 	\\
0.345 &	4.83	 &	3.17	         &	1.588	       & 0.009	         &	0.010 	  & 0.009 &	0.016 	\\
0.375 &	5.25	 &	3.10	         &	1.544	       & 0.013	         &	0.016 	  & 0.008 &	0.023 	\\
      &   &  &                     &                 &       &         &                \\
\hline\hline
\end{tabular}
\end{center}
\caption{\label{tabdp} The ratio of deuteron to proton DIS cross section at the selected $x$, $Q^2$, and $W$ kinematics of
MARATHON.  Listed are the statistical (stat), point-to-point (ptp), and overall/scale systematic (syst) components of the 
total (tot) error.  The latter is the quadrature sum of the three components.
The overall/scale systematic component of $\pm0.55\%$ is due to the uncertainties in the nominal gas
target densities of the hydrogen and deuterium gases (combined in quadrature).}
\end{table} 

\begin{table}
\begin{center}
\begin{tabular}{cccccccc}
\hline\hline
      &	 	&               & 	        &               &               &       &               \\
$x$ & $Q^2$ & $W$ & $\sigma_h / \sigma_t$ & $\Delta_{\rm stat}$ & $\Delta_{\rm ptp}$ & $\Delta_{\rm syst}$ & $\Delta_{\rm tot}$ \\
      &	$({\rm GeV}/{\it c})^2$ 	& ${\rm GeV}/{\it c}^2$	& 	&       &                &   &  \\
      &	 	&   & 	&       &                &   &  \\
\hline\hline
      &	 	&               & 	        &               &               &       &               \\
0.195 &	2.73	&	3.49	&	1.112	&	0.003	&	0.005	& 0.008	& 0.010 	\\
0.225 &	3.15	&	3.42	&	1.124	&	0.003	&	0.004	& 0.008	& 0.010 	\\
0.255 &	3.57	&	3.36	&	1.141	&	0.004	&	0.004	& 0.008	& 0.010 	\\
0.285 &	3.99	&	3.30	&	1.160	&	0.005	&	0.005	& 0.008	& 0.011 	\\
0.315 &	4.41	&	3.24	&	1.154	&	0.005	&	0.004	& 0.008	& 0.011 	\\
0.345 &	4.83	&	3.17	&	1.171	&	0.006	&	0.005	& 0.008	& 0.011 	\\
0.375 &	5.25	&	3.10	&	1.177	&	0.008	&	0.006	& 0.008	& 0.013 	\\
0.405 &	5.67	&	3.03	&	1.219	&	0.009	&	0.004	& 0.009	& 0.014 	\\
0.435 &	6.09	&	2.96	&	1.206	&	0.010	&	0.005	& 0.009	& 0.014 	\\
0.465 &	6.51	&	2.89	&	1.226	&	0.010	&	0.004	& 0.009	& 0.014 	\\
0.495 &	6.93	&	2.82	&	1.235	&	0.010	&	0.004	& 0.009	& 0.014 	\\
0.525 &	7.35	&	2.74	&	1.260	&	0.011	&	0.004	& 0.009	& 0.015 	\\
0.555 &	7.77	&	2.67	&	1.267	&	0.011	&	0.004	& 0.009	& 0.015 	\\
0.585 &	8.19	&	2.59	&	1.268	&	0.012	&	0.004	& 0.009	& 0.016 	\\
0.615 &	8.61	&	2.50	&	1.268	&	0.013	&	0.004	& 0.009	& 0.016 	\\
0.645 &	9.03	&	2.42	&	1.292	&	0.013	&	0.004	& 0.009	& 0.017 	\\
0.675 &	9.45	&	2.33	&	1.289	&	0.014	&	0.004	& 0.009	& 0.018 	\\
0.705 &	9.87	&	2.24	&	1.309	&	0.014	&	0.004	& 0.009	& 0.017 	\\
0.735 &	10.3	&	2.14	&	1.302	&	0.013	&	0.004	& 0.009	& 0.017 	\\
0.765 &	10.7	&	2.04	&	1.316	&	0.014	&	0.004	& 0.009	& 0.017 	\\
0.795 &	11.1	&	1.94	&	1.312	&	0.015	&	0.004	& 0.009	& 0.018 	\\
0.825 &	11.9	&	1.84	&	1.301	&	0.017	&	0.004	& 0.009	& 0.020 	\\
      &	 	&               & 	        &               &               &       &               \\
\hline\hline
\end{tabular}
\end{center}
\caption{\label{tabht} The helion to triton DIS cross section ratio (after normalization, see text) at the
$x$, $Q^2$, and $W$ kinematics of MARATHON.  
Listed also are the statistical (stat), point-to-point (ptp) and overall/scale systematic (syst) components
of the total (tot) error.  The latter is the quadrature of the three components.}
\end{table}   

\begin{table}
\begin{center}
\begin{tabular}{cccccccc}
\hline\hline
      &	 	&               & 	        &               &               &       &               \\
$x$	& ${\cal R}_{ht}$  &  $\Delta{\cal R}_{ht}$ 
& $F_2^n/F_2^p$  & $\Delta_{\rm stat}$ & $\Delta_{\rm ptp}$ & $\Delta_{\rm syst}$ & $\Delta_{\rm tot}$ \\
      &	 	&               & 	        &               &               &       &               \\
\hline\hline
      &	 	&               & 	        &               &               &       &               \\
0.195	&   0.9989 & 	0.0009 & 0.724	& 0.005 & 0.011	& 0.016 & 0.020	\\
0.225	&   0.9990 & 	0.0009 & 0.701	& 0.006 & 0.008	& 0.016 & 0.019	\\
0.255	&   0.9991 & 	0.0009 & 0.668	& 0.008 & 0.008	& 0.015 & 0.019	\\
0.285	&   0.9993 & 	0.0008 & 0.635	& 0.008 & 0.009	& 0.014 & 0.019	\\
0.315	&   0.9997 & 	0.0009 & 0.647	& 0.010 & 0.008	& 0.015 & 0.019	\\
0.345	&   1.0003 & 	0.0008 & 0.618	& 0.010 & 0.008	& 0.014 & 0.019	\\
0.375	&   1.0010 & 	0.0008 & 0.610	& 0.013 & 0.010	& 0.014 & 0.021	\\
0.405	&   1.0019 & 	0.0008 & 0.547	& 0.014 & 0.006	& 0.013 & 0.020	\\
0.435	&   1.0029 & 	0.0007 & 0.567	& 0.015 & 0.007	& 0.013 & 0.021	\\
0.465	&   1.0039 & 	0.0007 & 0.540	& 0.015 & 0.006	& 0.013 & 0.020	\\
0.495	&   1.0049 & 	0.0007 & 0.528	& 0.014 & 0.006	& 0.012 & 0.020	\\
0.525	&   1.0058 & 	0.0007 & 0.496	& 0.015 & 0.006	& 0.012 & 0.020	\\
0.555	&   1.0067 & 	0.0007 & 0.489	& 0.015 & 0.006	& 0.012 & 0.020	\\
0.585	&   1.0074 & 	0.0008 & 0.489	& 0.016 & 0.006	& 0.012 & 0.020	\\
0.615	&   1.0081 & 	0.0009 & 0.489	& 0.016 & 0.005	& 0.012 & 0.021	\\
0.645	&   1.0087 & 	0.0010 & 0.461	& 0.016 & 0.006	& 0.011 & 0.020	\\
0.675	&   1.0093 & 	0.0013 & 0.466	& 0.018 & 0.006	& 0.011 & 0.022	\\
0.705	&   1.0098 & 	0.0017 & 0.442	& 0.016 & 0.005	& 0.011 & 0.020	\\
0.735	&   1.0104 & 	0.0020 & 0.451	& 0.016 & 0.005	& 0.011 & 0.020	\\
0.765	&   1.0111 & 	0.0024 & 0.436	& 0.016 & 0.006	& 0.011 & 0.020	\\
0.795	&   1.0118 & 	0.0030 & 0.441	& 0.017 & 0.006	& 0.011 & 0.022	\\
0.825	&   1.0125 & 	0.0043 & 0.455	& 0.020 & 0.009	& 0.011 & 0.024	\\
      &	 	&               & 	        &               &               &       &               \\
\hline\hline
\end{tabular}
\end{center}
\caption{\label{tabnp} The $F_2^n/F_2^p$ ratio for the MARATHON $x$ kinematics.  Listed also are the
ratio's statistical (stat), point-to-point (ptp), and overall/scale (syst) components of the total (tot)
error.  The latter is the quadrature of the three components.  Also listed are the values for the
${\cal R}_{ht}$ super-ratio and their uncertainties used in the $F_2^2/F_2^p$ ratio extraction (see text).}
\end{table}

\end{document}